# Structural Characterization of Self-Assembled Monolayers of Organosilanes Chemically Bonded on Silica Wafers by Dynamical Force Microscopy

S. Navarre, F. Choplin, J. Bousbaa, B. Bennetau$^{\alpha}$

*Laboratoire de Chimie Organique et Organométallique, UMR 5802 CNRS, Université Bordeaux I, 351 cours de la Libération, 33405 Talence Cedex.*

L. Nony, J.-P. Aimé$^{\beta}$

*Centre de Physique Moléculaire Optique et Hertzienne, UMR 5798 CNRS, Université Bordeaux I, 351 cours de la Libération, 33405 Talence Cedex.*

**Abstract**

In this article, a dynamical force microscopy study of self-assembled monolayers of organosilanes, grafted on a silica support, is reported. Organosilanes, terminated either with a functional group (ethylene glycol) or with a methyl group, were used. The influence of the reaction time and the solvent composition on the grafting was investigated to improve the homogeneity of the self-assembled monolayers. Numerical simulations of approach-retract curves, obtained in the tapping mode, were performed and compared to experimental ones. Informations, such as mechanical response and height of the grafted organic layers, have been obtained.

---

$^{\alpha}$ Corresponding author. Tel.: 0033 5 56846275; Fax. : 0033 5 56846994 ; e-mail : b.bennetau@lcoo.u-bordeaux.fr

$^{\beta}$ Corresponding author. Tel.: 0033 5 56848956; Fax. : 0033 5 56846970 ; e-mail : jpaime@cribx1.u-bordeaux.fr

## Introduction

Since their discovery by Sagiv,[1] the grafting of long-chain organosilane compounds on hydroxylated solid surfaces to form dense self-assembled monolayers (SAM), is now widely used.[2] They have applications in a number of fields including lubrication, adhesion, and appear as being well-adapted substrates to study biological systems such as proteins[3] or DNA molecules.[4-6] Even though the mechanism of formation of the film and its quality remain dependent of a wide variety of parameters[7-10] (chain-length, temperature, solvent, reaction time, etc...), it's generally admitted that the use of long-chain *n*-alkyltrichlorosilanes favors the self assembling process producing a highly oriented film, representing a quasi-crystalline-like phase in which the hydrocarbon chains are perpendicularly oriented to the silica surface.

Various techniques have been used to investigate and characterize the self-assembled films[11] (FTIR, XPS, AFM, ellipsometry, contact angles measurements...). Recently in our group, unenhanced Raman spectroscopy was used to characterize the homogeneity of a SAM grafted on a $SiO_2$/Si non p-doped/Au surface at the micron scale. A mapping of the surface was obtained using a confocal micro-Raman spectrometer.[12]

Atomic force microscopy (AFM), as a local probe method, has been widely used to investigate the self-assembling properties of such layers.[13-15] It's expected that the SAM has a low elastic modulus. It should exhibit specific mechanical properties. Such properties are accessible with an AFM and in particular with a dynamical force microscope (DFM). Therefore, DFM provides an additional, useful, way to characterize the SAM properties. Among DFM, Tapping mode is now widely used. The first aim of this mode is to reduce the shear forces at the interface between the tip and the surface. Last few years, numerical and theoretical developments describing the non-linear behavior of the oscillating tip-cantilever system (OTCL) at proximity of the surface have been developed.[16-18] The non-linear behavior rises when the OTCL comes close to the surface, within the range of the nanometer. Taking advantage of the great sensitivity due to the non-linear dynamical properties of the OTCL, the Tapping mode has now a wide variety of applications in which soft materials can be investigated without significant damages.[19-23] Following that way, SAM appear as model samples to be investigated by DFM. Thus, information at a nanometric scale on homogeneity, chemical composition, thickness and mechanical response can be obtained.[24-26]

It is well-known that to investigate DNA properties with an AFM, or to have reproducible DNA chips, the main difficulty is the control of the chemical properties of the modified substrates. Commercially available silicon compounds, such as (3-aminopropyl)triethoxysilane (APTES) or (glycidoxypropyl)triethoxysilane (GPTES), were find to be inconsistent, by us or others,[27] for this purpose. A new series of silyl coupling agents, with different alkyl chain lengths, were synthesized.[28] These molecules, while being able to be the suitable ones for DNA chips applications,[29] exhibit defect structures with aggregates.

In this paper, in order to improve the quality of the SAM for DNA chips applications, and also to have suitable substrates to deposit DNA molecules, we chose to study by DFM the influence of parameters such as reaction time and solvent composition on the SAM formation. After a brief technical presentation, the different DFM images of SAM, obtained with two different silylated coupling agents (**1** and **2**, see fig. 1), are discussed then a paragraph is dedicated to the use of approach-retract curves as a simple and easy way to obtain informations such as mechanical response and height of the grafted organic layers.

**I - Materials and methods**

*I - 1- Formation of monolayers*

Monolayers were performed on bipolished silicon wafers (Micropolish) under cleanroom conditions (class 100). The substrates were cleaned by a wet chemical treatment in a freshly prepared 2% Hellmanex® solution with MilliQ deionized water solvent for two hours at room temperature (~20°C) followed by an extensive rinse with MilliQ deionized water. Under an atmosphere of argon, the substrates were dried[30] at 80°C for 40 min, cooled down at room temperature then immerged into a dilute solution of 2-(22-trichlorosilyl-docosoxy)-ethyl acetate (TSDOEA, **1**), or 22-trichlorosilyl-docosane (TSD, **2**), freshly prepared, in cyclohexane/chloroform (v/v; 90/10) or in cyclohexane/hexadecane/chloroform (v/v/v; 45/45/10). The synthesis of TSDOEA or TSD have been reported previously. The reaction solution was maintained at a constant temperature of 18°C ± 1°C, under an argon atmosphere, for 18 h. After removal from the reaction solution, the silanized wafers were washed with chloroform two times for 5 minutes, under ultrasonic condition. With the trichlorosilane **1**, the deprotection of hydroxyl groups was performed under basic conditions (0.5 M KOH; $H_2O$/EtOH (v/v; 1/1)).

### I - 2 - Atomic Force Microscopy

Experiments were performed with a Nanoscope III[31] into a glove box, under controlled atmosphere at the ppm of water and a few ten ppm of dioxygen. The controlled atmosphere prevents the hydration of the surface and is helpful to obtain reproducible and robust results. In such conditions, the behavior of the OTCL is very stable.

During a Tapping experiment, the tip-cantilever system is kept vibrating at a given drive frequency and a drive amplitude. During a scan, a feedback loop keeps constant a chosen oscillation amplitude (the setpoint) by varying the vertical position of the sample. The two, simultaneous, recorded images are the vertical displacements necessary to keep constant the amplitude −the so-called height image− and the corresponding phase values of the OTCL −the phase image−. To understand the origin of the contrast of the images, it's worth knowing changes of the amplitude and phase, A and $\varphi$ respectively, as functions of the OTCL-surface distance D. Recording A(D) and $\varphi$(D) is achieved by making approach-retract curves. Approach-retract curves are done at a given location (X;Y) in the horizontal plane of the sample, then a periodic motion along the vertical Z-axis is performed. In these experiments, approach-retract curves provide information on the properties of the dynamical behavior of the OTCL as a function of D and properties at the local scale at the (X;Y) location of the sample. The ultimate goal is to extract the sample properties from variations of A(D) and $\varphi$(D). For example, the slope of the variation of the amplitude as a function of the distance during an approach-retract curve is a fundamental parameter. Firstly, the slope contains information about the nanomechanical properties of the sample (see below) and secondly, the slope at a given setpoint, (amplitude reduction at which an image is recorded) controls the vertical displacement of the piezoelectric actuator.

It's beyond the scope of this paper to describe the non-linear behavior of the OTCL at proximity of the surface. Previous papers already dealt with this subject.[32, 33] For instance, one can readily know if whether or not the tip touches the surface. When the oscillator experiences a dominant attractive regime, the tip does not touch the surface while, when the oscillator behavior is dominated by a repulsive regime, intermittent contact situations occur. The transitions from a repulsive dominant regime to an attractive dominant one are obtained by varying the drive amplitude of the oscillator from large to small values. For the largest amplitudes, variations of the oscillation amplitudes are sensitive to the surface topography and local mechanical properties of the sample. In that case, the tip mainly experiences a pure repulsive regime and the contribution of the attractive interaction between the tip and the sample becomes negligible. The quality factor of the OTCL, Q, and the contact stiffness of the tip into the sample control the sensitivity. Under atmospheric conditions, typical quality factors of the OTCL used are about 400. In that case, the DFM can access to a maximal elastic modulus of the sample of about the GPa. For a modulus of the sample above that value, the mechanical response cannot be measured and the oscillator experiences a surface that behaves as a hard one. The slope of the amplitude variation as a function of the distance between the tip and the surface is about 1. That means that for 1 nanometer of vertical displacement of the sample, the amplitude is reduced by the equivalent quantity. For samples, whose modulus is smaller than that value, the mechanical response becomes accessible. The slope is smaller than 1. The image obtained during the scan becomes a mixing of both the topography and the mechanical response of the sample and therefore heights might not be the true ones. Nevertheless, even though it is not so well adapted, we will keep the expression of "height" image, as it's the one commonly used.

Thus, the contrast of the images can not be described without the help of the approach-retract curves and a modeling of the oscillating behavior of the OTCL.

Images and approach retract curves given in this work were recorded with an *ultrasharp* tip,[34] from which a small radius is expected, typically a few nanometers. The size of the tip was checked by performing approach retract curves. With that tip, a pure attractive regime was never observed, even at oscillation amplitudes as small as 5 nm. This result clearly indicates that the size of the tip is smaller than those given by the usual Nanosensors cantilevers.[32, 33] While the use of *ultrasharp* tip requires to be cautious to save the tip shape, study of soft materials reduces chance of destroying the tip apex. The use of a small tip size has two objectives : on one hand we want to minimize the attractive interaction in order to uniquely have access to mechanical properties of the fluid-like behavior of the SAM and on the other hand the access to those mechanical properties requires a small contact area between the tip and the monolayer. The latter constraint can be understood as follow : with a simple geometrical description of the contact between a sphere and a plane surface, one gets a contact diameter $\phi$ scaling as $\phi \propto 2\sqrt{R\delta}$, where R is the tip radius and $\delta$ the tip indentation. Moreover, the contact stiffness $k_s$, e.g. the sample stiffness probed by the cantilever is given by $k_s \propto E_s\phi$, where $E_s$ is the elastic modulus. Therefore a large tip radius increases the local stiffness and in turn lowers the sensitivity of the measurement. To ensure that the tip experiences a pure repulsive regime, a large oscillation amplitude was used, typically 50 nm for the free amplitude (see below). We focus the discussion on the height images, the phase being used to ensure that the intermittent contact situation is reached.

**II - Images of the SAM**

*II - 1 - With trichlorosilane 1 (TSDOEA), after deprotection of the hydroxyl groups :*

The first sample investigated was made with the trichlorosilane **1**, this coupling agent being dedicated to DNA chips. The DFM image, given in figure 2, was performed on $5\times5\mu m^2$, the vertical contrast being 10 nm. The other experimental conditions are given in the caption. A dense layer of molecules is observed over the whole area with some ungrafted zones. Thus one have access to the silica substrate providing the opportunity to evaluate the layer thickness (3.7 nm; fig. 2b). This result is in good agreement with the expected height of the silane **1** (3.8 nm). This result indicates that only one monolayer is grafted on the silica. The layer homogeneity may be locally good, but nevertheless, some aggregates remain. Their sizes and shapes are different from a zone to another one and their heights are sometimes close to the one of the layer, e.g. 3.7 nm. Thus, it seems possible that they correspond to nucleation points for the growth of a new layer above the first one. Such aggregates are particularly disturbing for DFM analysis for many reasons : the first one is that a precise analysis requires a good description of the tip-surface interaction. The surface has to be locally flat with a very small roughness such that it may be assimilated to a plane. While for DNA chips we have shown that the presence of a few aggregates does not matter, in the case of studies of DNA molecules deposited onto a substratum, the aggregates may act as a significant perturbation.[6, 35] In that case, the DNA molecule can not be investigated without involving their contribution and so a precise measurement becomes impossible. The second reason is that aggregates can pollute the tip. A reasonable hypothesis would be that some molecules into the aggregates are not firmly fixed and could be adsorbed on the tip during the scan.

These elements led us to consider simpler structure of the silanes molecules, in particular without the ethylene-glycol group; therefore we used the silane **2**. In order to better control and enhance the homogeneity of the layers, the effect of two experimental parameters, the solvent composition and the reaction time, have been investigated.

*II - 2 -With trichlorosilane **2** (TSD):*

To make the results qualitatively easy to compare, the images obtained for each sample are shown on two different areas, $5\times5\mu m^2$ and $1\times1\mu m^2$ with an identical vertical contrast. Moreover, for each sample, several zones have been investigated. Thus a qualitative statistical distribution of the layer properties such as the density of aggregates, mechanical response, or structural defaults is obtained. The images shown are the most representative of the responses observed.

It is known that the use of long-chain *n*-alkyltrichlorosilanes, such as **2**, provides dense and robust SAM. The grafting of TSD was achieved under the same experimental conditions than the ones used for TSDOEA.

The height images obtained are given in the figures 3a and 3b. The structure of the layer appears to be different than the one observed with molecules (**1**). In particular, no ungrafted zones can be revealed on the areas investigated. Thus the layer's height is not so directly accessible than in the previous case. Fortunately, approach-retract curves (see section III) unambiguously show that only one monolayer is also grafted on the silica substrate. In this part, we only discuss the contribution of the solvent composition and reaction time to the homogeneity of the layer. On the two images, the presence of aggregates is observed. Their average height is around 0.5 nm. The only possibilities for their presence after the different cleaning treatments (see section I) could be either a head to tail pre-organization of the SAM in solution or a slow diffusion of the alkyl chains between the alkyl chains of the grafted SAM.

To reduce the density of aggregates, the solvent composition was modified. We decided to use hexadecane as co-solvent with the mixture cyclohexane/chloroform. Such a co-solvent should limit the head to tail penetration of 22-trichlorosilyl docosane (or 22-trihydroxylsilyl docosane after hydrolysis of the trichlorosilyl group) and could fill ungrafted areas during the growth of the covering. The grafting reaction was performed in the same condition than previously (time, concentration and temperature) but with cyclohexane/hexadecane/chloroforme 45/45/10 as solvent mixture. The images presented in figures 4, obtained in the same conditions than in figure 3, show the presence of aggregates but their number seems to be reduced. Another point is that the images shows a different structure of the layer. This experimental observation could be due to a better organization of the SAM. Nevertheless, the density of aggregates is still large.

In a recent paper, Saavedra et al. confirmed the importance of reaction time in the covering of silicon wafers with alkyltrichlorsilane bearing thioacetate and/or acetate tail groups. A reaction time of about one or two hours (depending on the alkyltrichlorosilane) is efficient to cover the surface. AFM study clearly show that if the silica is "quickly" grafted, the quality of the SAM is strongly dependent of the reaction time. Longer reaction time induces the presence of aggregates. To explain that, the authors suggested a reaction of the thioacetate or acetate function with free silanol of preformed aggregates in solution. Depending on the nature of the surface, the grafting time is ranging in a few minutes to a few hours to obtain a good covering. In our case, such a reaction can not occur but the observed aggregates are strongly adsorbed at the surface of the monolayer. This is probably due to a head to tail trapping of the alkyl part of the coupling agent. The silanol function at the surface can then react with other silanol present in solution and create aggregates. Reaction time is then an key parameter to obtain a completely formed monolayer.

Therefore, according to the results of Saavedra et al., the reaction time was reduced from 18 h to 2 h. The figure 5 shows that, within the same experimental conditions, only a few aggregates are observed. The SAM is homogeneous on a large scale (5×5 µm$^2$). In that conditions of grafting, a homogeneous SAM at the micrometer scale is produced.

To compare the effect of the solvent and the effect of the reaction time on the density of aggregates, the grafting was realized with the first solvent mixture cyclohexane/chloroform 90/10 and a reaction time of 2 hours (results not shown); the other parameters were identical. In that conditions, no decrease of the quantity of aggregates was noticed. This result seems to show that the solvent has a great influence on the homogeneity of the grafting. The pre-organization of the alkyltrichlorosilane in solution could have an important role for the homogeneity of a SAM. Indeed it is known that the organic layer is grafted by pre-formed and pre-organized "islands" of coupling agent in solution. Hexadecane as co-solvent should favor the pre-organization in solution of the film and could fill the empty place as it was already suggested with alkanethiol compounds deposited on gold.[36]

It has been already pointed out that the experimental conditions of the grafting must be optimized for each trichlorosilane. However, in our case, the most efficient experimental conditions (co-solvent and reaction times) with the silane **2**, did not lead to significant improvement of the quality of the SAM with the silane **1**.

### III - SAM's properties investigated with approach-retract curves

*III - 1 - Thickness measurement with approach-retract curves*

The approach-retract curves provide the opportunity to link the evolution of the amplitude and the phase of the oscillator as a function of the tip-sample distance D. In the case of the SAM, the use of large amplitudes to perform the curve gives the information of the indentation depth of the tip through the layer before it touches the hard silica substrate without involving the contribution of the attractive interaction between the tip and the surface. The layer thickness may therefore be obtained. Thus the expected shape of the curve is, firstly a low amplitude reduction, with a slope smaller than 1, on a vertical distance which is characteristic of the layer thickness and secondly a larger amplitude reduction, with a slope equal to 1, meaning that the tip is into contact with the hard substrate. Thus, the thickness of the SAM is evaluated by measuring the vertical gap between the location of the surface and the one for which the slope of the intermittent contact is 1. It's beyond the scope of this paper to discuss the precise positioning of the surface but, for the intermittent contact situations, it can be estimated by the point at which the bifurcation occurs[37] (see figs. 7). This argument is particularly valid when large amplitudes are used. The previous reference shows that the error made on the surface positioning is about 0.1 nm. Nevertheless, even though large amplitudes are used, the mechanical response of the layer is only accessible if the size of the tip is small enough (see part I).

Since no ungrafted zones were observed for the samples made with the trichlorosilane **2**, the thickness of the layer was indirectly deduced from the approach-curves. The figures 6a and 6b give an experimental curve in amplitude and phase respectively recorded on the sample surface shown in figure 5. The experimental conditions are given in the caption. After the instability, the amplitude is weakly reduced with a slope close to 1%. The slope close to 1 is only reached after a vertical displacement of the surface of about 3.3 nm. Moreover, the phase, above -90°, unambiguously shows that the tip always experiences an intermittent contact situation, thus allows the use of the above method to evaluate the thickness of the SAM.

The calculated height of a SAM, of 22-trichlorosilyl docosane, grafted on a silica wafer with an all trans conformation is of about 3.4 nm. The expected theoretical value and the one obtained by the approach-retract curve are nearly identical. Therefore, if no defect or hole is available the experimental approach-retract curves can be used to measure the height of a soft material grafted on a hard support with a very good precision.

*III - 2 - Numerical simulation of an approach-retract curve*

To go a step further and have an evaluation of the elastic modulus of the layer, numerical simulations of an approach-retract curve on a model SAM were developed. In the simulation, the SAM is characterized by a thickness $\xi$, an elastic modulus E and an Hamaker constant H. The substrate parameters are its modulus $E_s$ and Hamaker constant $H_s$. The numerical simulation solves the usual non-linear second order differential equation based on the forced, damped harmonic oscillator plus an interacting term with a Runge-Kutta 4 method[38] from arbitrary initial conditions :

$$\ddot{z}(t) + \frac{\omega_0}{Q}\dot{z}(t) + \omega_0^2 z(t) = \frac{f}{m}\cos(\omega t) + \frac{f_{int}}{m}, \qquad (1)$$

z(t) is the instantaneous position of the tip; m, $\omega_0$ and Q are the mass, the resonance frequency and the quality factor respectively. f and $\omega$ are the drive force and drive frequency. $f_{int}$ is the interacting force between the tip and the surface. An adiabatic criterion ensures that the simulation calculates an harmonic stationary state at each implementation step. Amplitude and phase variations are then calculated thanks to a numerical lock in amplifier filtering the first harmonic. With D the tip-sample distance for the oscillator at rest, non-contact situations (z(t) - D < 0) and intermittent contact situations either into the SAM (z(t) - D > 0), or into the silica (z(t) - ($\xi$ + D) >0) are investigated during an oscillation period of the tip. The attractive force is a Van der Waals disperse force between a sphere and a plane surface whereas the Hertz model[39] is used for the repulsive part :

$$\begin{cases} f_{int} = \frac{4}{3}\sqrt{R}E_I\delta^{3/2}, \forall z(t) > D \\ f_{int} = -\frac{H_I R}{6[D - z(t)]^2}, \forall z(t) < D \end{cases}, \qquad (2)$$

where R, $E_I$, $H_I$ are respectively the tip's apex radius, elastic's modulus and Hamaker constant of the SAM or of the substrate. $\delta$ = z(t) - D is the indentation depth of the tip into the sample. The contact between the tip and the SAM occurs when z(t) = D + $d_c$. $d_c$ is a contact distance whose value is 0.165 nm for most of the organic materials.[40] If $\delta < \xi$, the tip uniquely experiences the SAM, else it experiences both the SAM and the wafer. The description of the SAM assumes a homogeneous mechanical response of the layer for any indentation depth smaller than that of the thickness $\xi$. Beyond that value, the tip touches the hard surface leading to a elastic modulus discontinuity.

The figures 7 show the numerical simulation of an approach-retract curves obtained on a SAM whose thickness is 3.3 nm, in agreement with the height measured from the experimental curves. To get a good qualitative agreement with the experimental variations, two values of the elastic modulus are shown : 50 Mpa and 20 Mpa. Above the former value, the amplitude is reduced too quickly as a function of the vertical displacement and below the latter one, the amplitude variation does not exhibit a plateau-like shape.[41] The numerical oscillation conditions are the same than the experimental ones. The phase has also a behavior close to the observed experimental one. Therefore, the experimental approach-retract curve confirms the presence of only one soft organic layer grafted on a hard support with an elastic modulus of a few ten MPa.

**Conclusion**

Atomic Force Microscopy, in the Tapping mode, was used to study the influence of experimental conditions such as reaction time and solvent mixture composition on the quality of a SAM. An homogeneous SAM was obtained with 22-trichlorosilyl docosane (good covering of the surface without aggregate) by varying the experimental conditions. The influence of solvent and reaction time was pointed out to improve the monolayer formation with the alkylated trichlorosilane, but not with the terminally functionalized (ethylene-glycol group) trichlorosilane. Approach-retract curves were used to measure the mechanical response of the layer and its thickness. With the help of a numerical simulation of the SAM mechanical properties, a good agreement is obtained between the simulations and the experimental data.

# Captions

**Figure 1:**

Trichlorosilanes used for the formation of SAM.

**Figure 2 :**

Height image of 5×5 µm$^2$ (a) and cross section (b) of the SAM obtained with the trichlorosilane **1**, after deprotection of hydroxyl groups. The hole allows the measure of the thickness of the SAM. The value measured is of about 3.7 nm, in agreement with the expected height for a monolayer (3.8 nm, see text).

**Figure 3:**

Height images of 5×5 µm$^2$ (a) and 1×1 µm$^2$ (b) of the SAM obtained with the trichlorosilane **2,** grafted during 18 h with a solvent mixture cyclohexane/chloroform. The vertical contrast is 10 nm. The experimental conditions are $A_0$=78 nm, $\nu_0$=148.480 kHz. The quality factor of the oscillator is Q=380. The drive frequency is $\nu_{exc}$=148.270 kHz. The free amplitude and free phase are respectively equal to 54.5 nm and –43°. The setpoint amplitude is 53 nm.

**Figure 4:**

Height images of 5×5 µm$^2$ (a) and 1×1 µm$^2$ (b) of the SAM obtained with the trichlorosilane **2** grafted during 18 h with a solvent mixture cyclohexane/hexadecane/chloroform 45/45/10. The vertical contrast is 10 nm. The experimental conditions that have changed regardless to fig. 3 are $A_0$=119 nm and the free amplitude equal to 81.5 nm. The setpoint amplitude is 80 nm.

**Figure 5:**

Height images of 5×5 µm² (a) and 1×1 µm² (b) of the SAM obtained with the trichlorosilane **2** grafted during 2 h with a solvent mixture cyclohexane/hexadecane/chloroform 45/45/10. The vertical contrast is 10 nm. The experimental conditions which have changed regardless to fig. 3 are $A_0$=157 nm and the free amplitude, 109 nm. The setpoint amplitude is 108 nm.

**Figure 6:**

Experimental approach-retract curve, amplitude (a) and phase (b) obtained on the sample of the figure 5. Concerning the amplitude, the approach part of the curve is indicated by the arrow up and the retract by the arrow down. The experimental conditions are the same than fig. 3. The thickness of the layer deduced as explained in text is of about 3.3 nm, in agreement with the expected one : 3.4 nm.

**Figure 7:**

Numerical approach-retract curve, amplitude (a) and phase (b). The numerical parameters of the oscillator are the same than the ones given in fig. 3. The other numerical parameters are R=1 nm, $H_{SAM}$=3.10$^{-20}$ J and 10$^{-20}$ J, $E_{SAM}$=20 MPa and 50 MPa (gray and black lines), ξ=3.3 nm, and $H_S$=10$^{-20}$ J and $E_S$=100 GPa in both cases.

Figures

Figure 1:

Cl$_3$Si—(CH$_2$)$_n$—O—CH$_2$CH$_2$—OAc     Cl$_3$Si—(CH$_2$)$_n$—CH$_3$

         **1**                                  **2**

n = 22

S. Navarre et al.

**Structural Characterization of Self-Assembled Monolayers of Organosilanes Chemically Bonded on Silica Wafers by Dynamical Force Microscopy**

Figure 2:

**(a)**

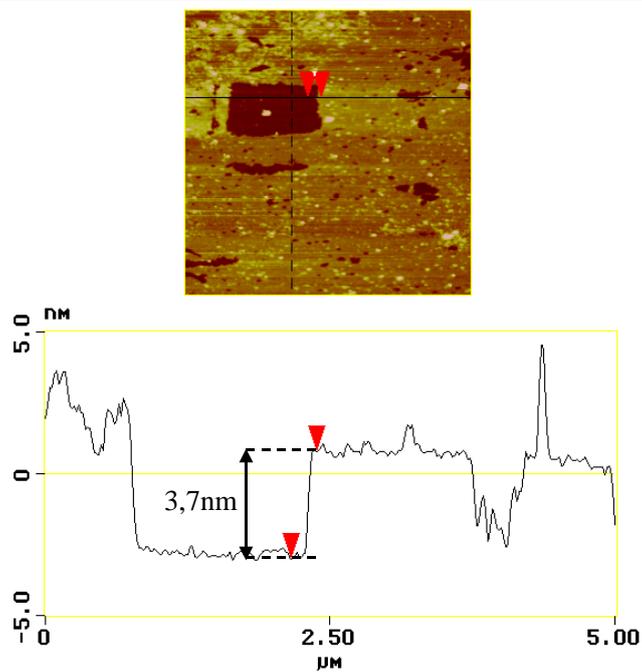

**(b)**

S. Navarre et al.

**Structural Characterization of Self-Assembled Monolayers of Organosilanes Chemically Bonded on Silica Wafers by Dynamical Force Microscopy**

Figure 3:

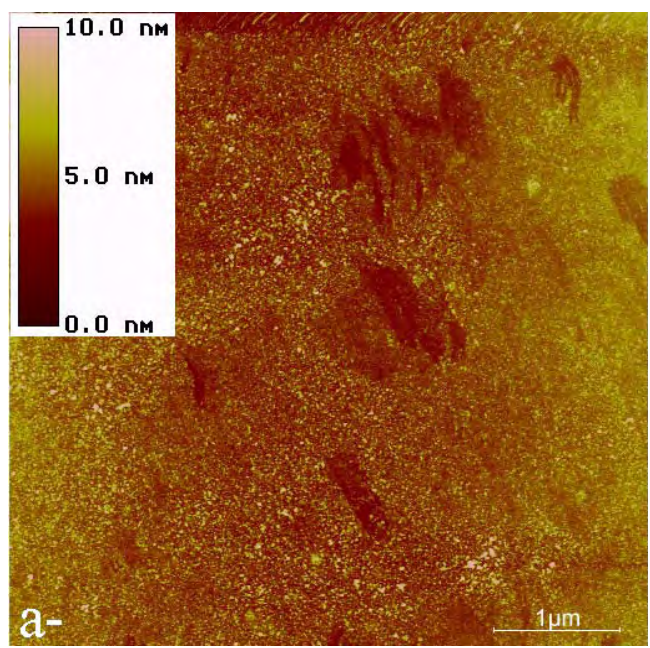

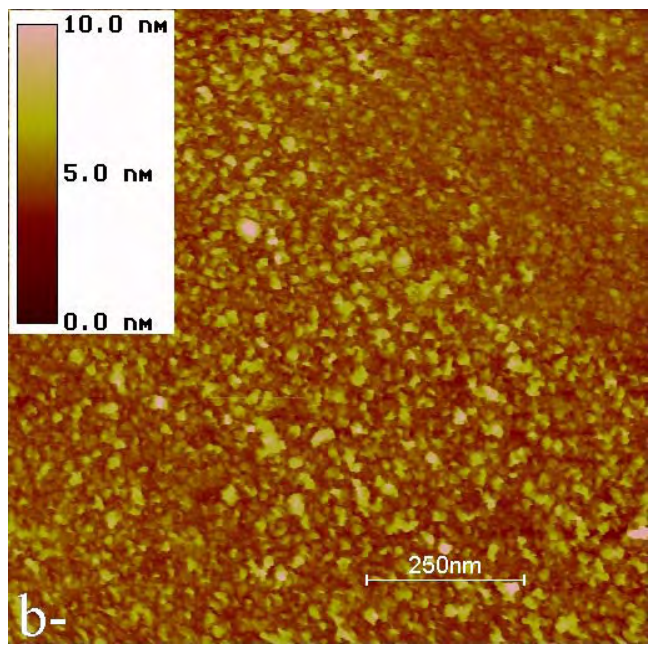

S. Navarre et al.

**Structural Characterization of Self-Assembled Monolayers of Organosilanes Chemically Bonded on Silica Wafers by Dynamical Force Microscopy**

Figure 4:

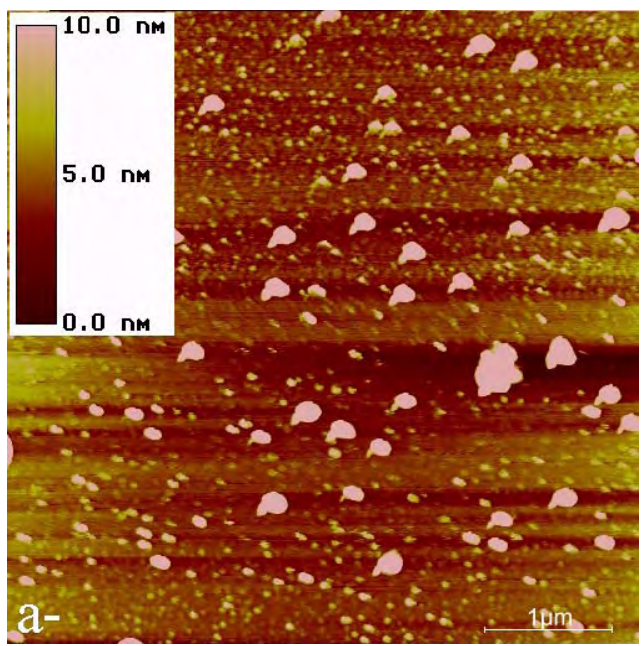

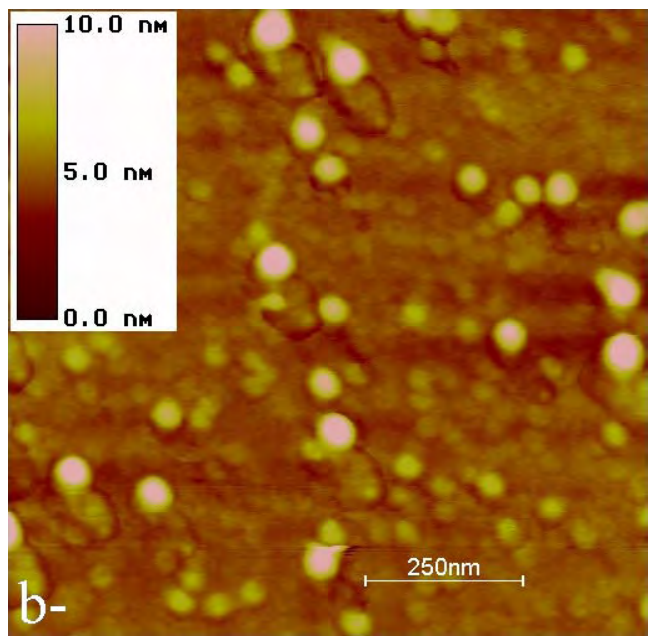

S. Navarre et al.

**Structural Characterization of Self-Assembled Monolayers of Organosilanes Chemically Bonded on Silica Wafers by Dynamical Force Microscopy**

Figure 5:

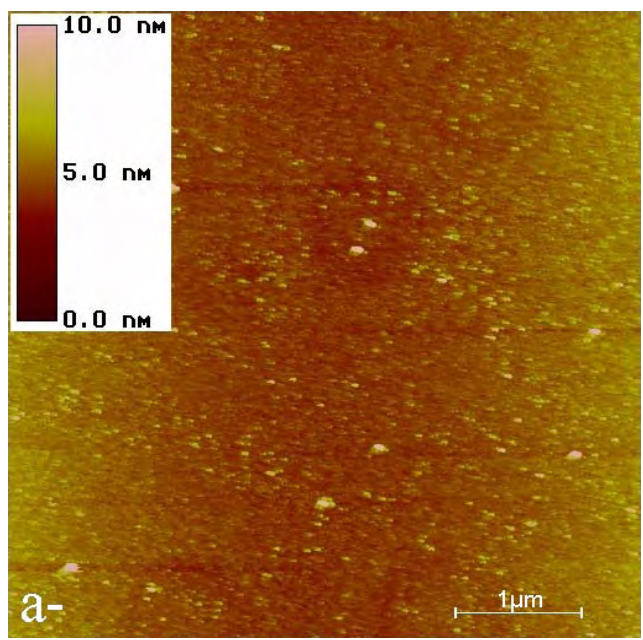

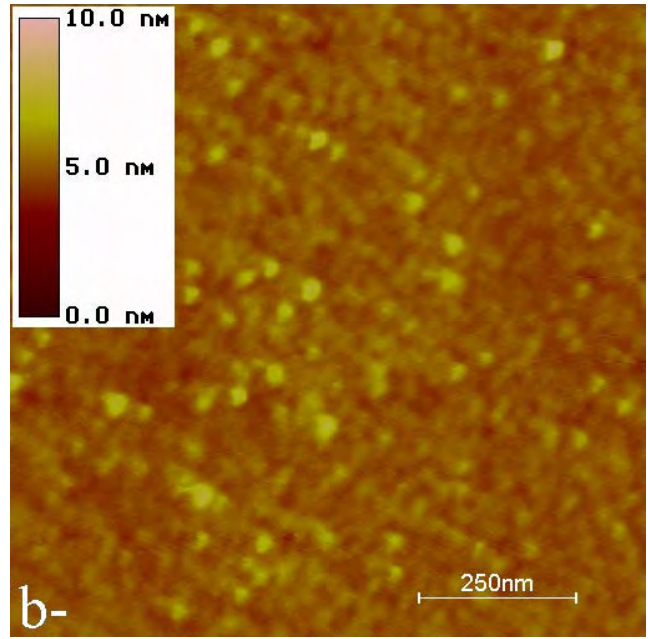

S. Navarre et al.

**Structural Characterization of Self-Assembled Monolayers of Organosilanes Chemically Bonded on Silica Wafers by Dynamical Force Microscopy**

**Figure 6:**

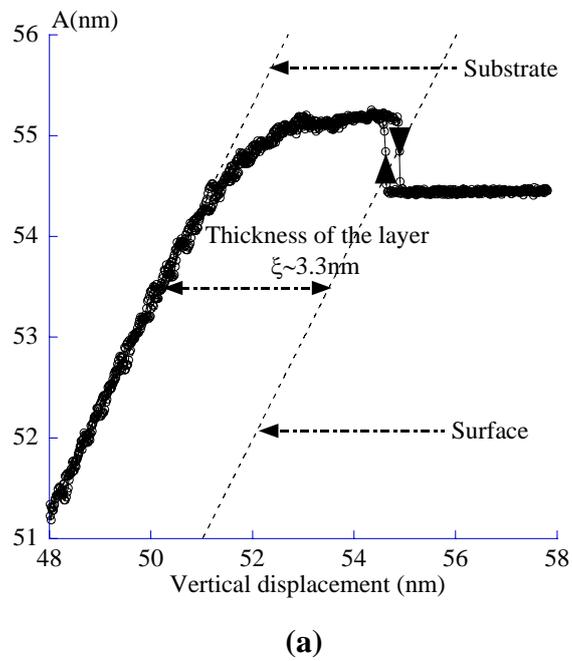

(a)

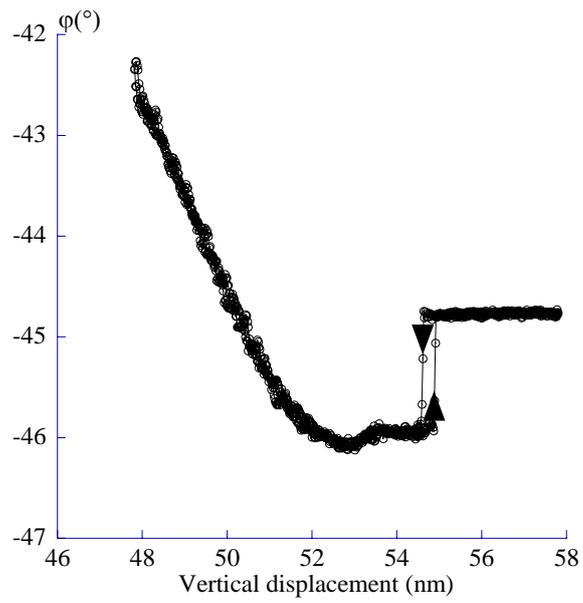

**(b)**

S. Navarre et al.

**Structural Characterization of Self-Assembled Monolayers of Organosilanes Chemically Bonded on Silica Wafers by Dynamical Force Microscopy**

**Figure 7:**

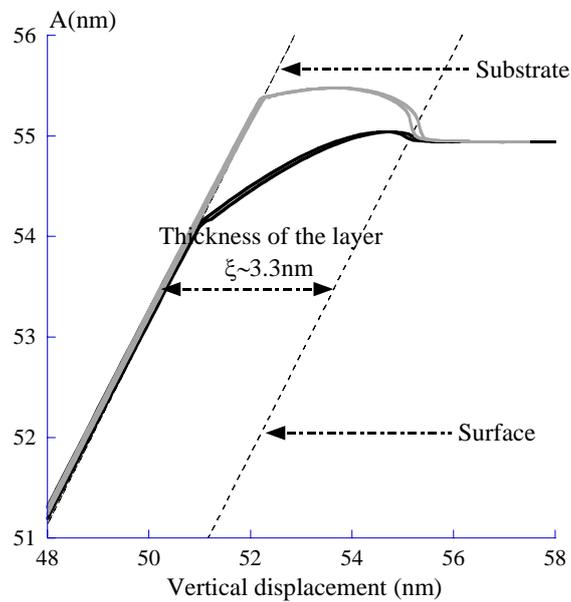

**(a)**

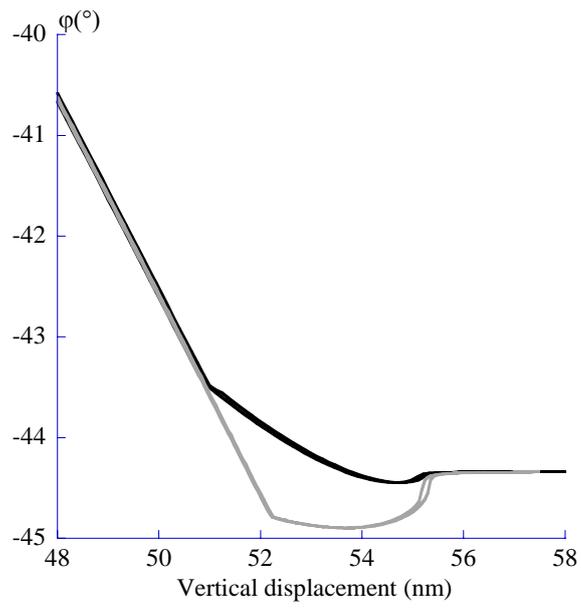

**(b)**

S. Navarre et al.

**Structural Characterization of Self-Assembled Monolayers of Organosilanes Chemically Bonded on Silica Wafers by Dynamical Force Microscopy**